\documentclass[reprint,superscriptaddress,amsmath,amssymb,aps,prc,twocolumn,nofootinbib]{revtex4-2}

\usepackage{graphicx}
\usepackage{hyperref}
\usepackage{color}
\usepackage[english]{babel}
\usepackage{bm}% bold math
\usepackage[T1]{fontenc}
\usepackage{threeparttable}
\usepackage{multirow}

\usepackage{lineno,hyperref}
\usepackage[output-decimal-marker={.},exponent-product=\cdot]{siunitx}

\begin{document}

\title{High-precision Penning-trap mass measurements of Cd and In isotopes at JYFLTRAP remove the fluctuations in the two-neutron separation energies}

%% Authors
\author{A.~Jaries}
\email{arthur.a.jaries@jyu.fi}
\affiliation{University of Jyvaskyla, Department of Physics, Accelerator laboratory, P.O. Box 35(YFL) FI-40014 University of Jyvaskyla, Finland}
\author{M.~Stryjczyk}
\email{marek.m.stryjczyk@jyu.fi}
\affiliation{University of Jyvaskyla, Department of Physics, Accelerator laboratory, P.O. Box 35(YFL) FI-40014 University of Jyvaskyla, Finland}
\author{A.~Kankainen}
\email{anu.kankainen@jyu.fi}
\affiliation{University of Jyvaskyla, Department of Physics, Accelerator laboratory, P.O. Box 35(YFL) FI-40014 University of Jyvaskyla, Finland}
\author{L.~Al~Ayoubi}
\affiliation{University of Jyvaskyla, Department of Physics, Accelerator laboratory, P.O. Box 35(YFL) FI-40014 University of Jyvaskyla, Finland}
\affiliation{Universit\'e Paris Saclay, CNRS/IN2P3, IJCLab, 91405 Orsay, France}
\author{O.~Beliuskina}
\affiliation{University of Jyvaskyla, Department of Physics, Accelerator laboratory, P.O. Box 35(YFL) FI-40014 University of Jyvaskyla, Finland}
\author{P.~Delahaye}
\affiliation{GANIL, CEA/DSM-CNRS/IN2P3, Boulevard Henri Becquerel, 14000 Caen, France}
\author{T.~Eronen}
\affiliation{University of Jyvaskyla, Department of Physics, Accelerator laboratory, P.O. Box 35(YFL) FI-40014 University of Jyvaskyla, Finland}
\author{M.~Flayol}
\affiliation{Universit\'e de Bordeaux, CNRS/IN2P3, LP2I Bordeaux, UMR 5797, F-33170 Gradignan, France}
\author{Z.~Ge}
\affiliation{GSI Helmholtzzentrum f\"ur Schwerionenforschung, 64291 Darmstadt, Germany}
\affiliation{University of Jyvaskyla, Department of Physics, Accelerator laboratory, P.O. Box 35(YFL) FI-40014 University of Jyvaskyla, Finland}
\author{W.~Gins}
\affiliation{University of Jyvaskyla, Department of Physics, Accelerator laboratory, P.O. Box 35(YFL) FI-40014 University of Jyvaskyla, Finland}
\author{M.~Hukkanen}
\affiliation{University of Jyvaskyla, Department of Physics, Accelerator laboratory, P.O. Box 35(YFL) FI-40014 University of Jyvaskyla, Finland}
\affiliation{Universit\'e de Bordeaux, CNRS/IN2P3, LP2I Bordeaux, UMR 5797, F-33170 Gradignan, France}
\author{D.~Kahl}
\affiliation{Extreme Light Infrastructure – Nuclear Physics, Horia Hulubei National Institute for R\&D in Physics and Nuclear Engineering (IFIN-HH), 077125 Bucharest-M\u{a}gurele, Romania}
\author{S.~Kujanp\"a\"a}
\affiliation{University of Jyvaskyla, Department of Physics, Accelerator laboratory, P.O. Box 35(YFL) FI-40014 University of Jyvaskyla, Finland}
\author{D.~Kumar}
\affiliation{GSI Helmholtzzentrum f\"ur Schwerionenforschung, 64291 Darmstadt, Germany}
\author{I.D.~Moore}
\affiliation{University of Jyvaskyla, Department of Physics, Accelerator laboratory, P.O. Box 35(YFL) FI-40014 University of Jyvaskyla, Finland}
\author{M.~Mougeot}
\affiliation{University of Jyvaskyla, Department of Physics, Accelerator laboratory, P.O. Box 35(YFL) FI-40014 University of Jyvaskyla, Finland}
\author{D.A.~Nesterenko}
\affiliation{University of Jyvaskyla, Department of Physics, Accelerator laboratory, P.O. Box 35(YFL) FI-40014 University of Jyvaskyla, Finland}
\author{S.~Nikas}
\affiliation{University of Jyvaskyla, Department of Physics, Accelerator laboratory, P.O. Box 35(YFL) FI-40014 University of Jyvaskyla, Finland}
\author{H.~Penttil\"a}
\affiliation{University of Jyvaskyla, Department of Physics, Accelerator laboratory, P.O. Box 35(YFL) FI-40014 University of Jyvaskyla, Finland}
\author{D.~Pitman-Weymouth}
\affiliation{Department of Physics and Astronomy, University of Manchester, Manchester M13 9PL, United Kingdom}
\author{I.~Pohjalainen}
\affiliation{University of Jyvaskyla, Department of Physics, Accelerator laboratory, P.O. Box 35(YFL) FI-40014 University of Jyvaskyla, Finland}
\author{A.~Raggio}
\affiliation{University of Jyvaskyla, Department of Physics, Accelerator laboratory, P.O. Box 35(YFL) FI-40014 University of Jyvaskyla, Finland}
\author{W.~Rattanasakuldilok}
\affiliation{University of Jyvaskyla, Department of Physics, Accelerator laboratory, P.O. Box 35(YFL) FI-40014 University of Jyvaskyla, Finland}
\author{A.~de Roubin}
\altaffiliation[Present address: ]{KU Leuven, Instituut voor Kern- en Stralingsfysica, B-3001 Leuven, Belgium}
\affiliation{Universit\'e de Bordeaux, CNRS/IN2P3, LP2I Bordeaux, UMR 5797, F-33170 Gradignan, France}
\author{J.~Ruotsalainen}
\affiliation{University of Jyvaskyla, Department of Physics, Accelerator laboratory, P.O. Box 35(YFL) FI-40014 University of Jyvaskyla, Finland}
\author{V.~Virtanen}
\affiliation{University of Jyvaskyla, Department of Physics, Accelerator laboratory, P.O. Box 35(YFL) FI-40014 University of Jyvaskyla, Finland}

\date{\today}% It is always \today, today,
             %  but any date may be explicitly specified

\begin{abstract}
We report on the first direct mass measurements of the $^{118,119}$Cd and $^{117-119}$In isotopes performed at the Ion Guide Isotope Separator On-Line facility using the JYFLTRAP double Penning trap mass spectrometer. The masses of $^{117}$In and $^{118}$Cd isotopes are in agreement with the literature, while $^{118,119}$In and $^{119}$Cd differ from literature by 49, 13 and~85 keV (6.1, 1.9 and 2.1 standard deviations), respectively. The excitation energy of the $^{118}$In first isomeric state, $E_x = 40.3(25)$~keV, was determined for the first time. The updated mass values removed the fluctuations observed in the two-neutron separation energies and lead to a smoother linear decrease of both isotopic chains. The $\log(ft)$ value for the $^{118}$Cd decay is also found to increase from 3.93(6) to 4.089(8). The reported results indicate an absence of significant structural changes around $N=70$.
\end{abstract}

\maketitle

% \linenumbers

\section{Introduction}

One of the most fundamental properties of a nucleus is its mass or in other words its binding energy. It contains a sum of all interactions between constituent protons and neutrons. An analysis of mass trends throughout isotopic chains, e.g. two-neutron separation energies, provides key information regarding changes in nuclear structure of the ground state. For example, shell closures appear as a steep decrease in the two-neutron separation energies after a magic neutron number has been crossed while onset of deformation manifest themselves as kinks in the otherwise smooth, linear decrease of the two-neutron separation energies (see e.g. Refs.~\cite{Lunney2003,Eronen2016,Garrett2022}). Cadmium ($Z=48$) and indium ($Z=49$) isotopic chains exhibit small deviations to the otherwise linear trend in two-neutron separation energies \cite{AME2020}. Atomic masses are also important for decay spectroscopy studies. The $\beta$-decay energy window $Q_\beta$, which is a mass difference between the parent and daughter nuclides, is a key ingredient to calculate $\log(ft)$ values \cite{Turkat2023}. 

While decay-energy measurements used to be important in the determination of atomic masses \cite{Huang2021}, nowadays more precise and accurate methods, such as Penning-Trap Mass Spectrometry (PTMS), are used. PTMS studies performed on nuclei away from stability indicated that many $\beta$-decay measurements had underestimated $Q_\beta$ values \cite{Audi2012}. This problem was associated with complex decay schemes and the pandemonium effect \cite{Audi2012}. The problem with the $\beta$-decay reliability was also observed around the valley of stability, with a striking example of a pair of stable nuclei, $^{102}$Pd and $^{102}$Ru. Their $Q_{\beta\beta}$ value differs by $10$ standard deviations ($10\sigma$) between the SHIPTRAP PTMS study \cite{Goncharov2011} and the decay and reaction studies \cite{Audi2012}. An independent JYFLTRAP PTMS measurement \cite{Nesterenko2019}, in agreement with SHIPTRAP, proves the reliability of the Penning trap measurements.

The masses of neutron-rich isotopes of $^{120-132}$Cd and $^{120-134}$In have been studied with ISOLTRAP at CERN \cite{Breitenfeldt2010,Atanasov2015,Manea2020}, JYFLTRAP at the Ion Guide Isotope Separator On-Line (IGISOL) facility \cite{Hakala2012,Kankainen2013,Nesterenko2020,Ruotsalainen2023}, Canadian Penning Trap (CPT) at Argonne National Laboratory \cite{VanSchelt2013,Orford2018} and TITAN at TRIUMF \cite{Lascar2017,Babcock2018,Izzo2021}. Nonetheless, the masses of the less exotic species are only known from transfer-reaction \cite{Hinds1967,Weiffenbach1971,Pilt1985} and $\beta$-decay \cite{McGinnis1955,Aleklett1982} studies. Thus, high-precision mass measurements are required to determine whether the deviations in mass trends are real effects arising from nuclear structure effects or due to potentially inaccurate mass values in literature. It should be also noted that the excitation energies of the isomeric states in $^{118}$In are currently unknown as the available data were deemed unreliable by the evaluators of the Atomic Mass Evaluation 2020 (AME20) \cite{Huang2021}. 

In this work we report on the first mass measurements of the neutron-rich $^{118,119}$Cd and $^{117-119}$In isotopes performed using the JYFLTRAP double Penning trap. The influence of the newly obtained mass values on binding-energy trends and $\log(ft)$ values is discussed.

\section{Experimental method}

The radioactive species were measured in three independent experiments performed at the IGISOL facility at the University of Jyv\"askyl\"a. The $^{118,119}$Cd and $^{119}$In isotopes were produced in proton-induced fission using a 25-MeV proton beam, delivered by the K130 cyclotron, with a 15-mg/cm$^{2}$-thick $^{nat}$U target for the cadmium isotopes and a 15-mg/cm$^{2}$-thick $^{232}$Th target for $^{119}$In. The $^{117,118}$In isotopes were produced in a fusion-evaporation reaction of a 50-MeV $\alpha$ beam with a 2.2~mg/cm$^{2}$-thick $^{nat}$Cd target. In all three experiments the reaction products were stopped in a helium-filled gas cell operating at about 300~mbar during the fission runs and about 250~mbar during the fusion-evaporation run. From the gas cell, the ions were extracted, guided through a sextupole ion guide \cite{Karvonen2008}, accelerated to 30$q$~kV and separated with respect to their mass-to-charge ratio $A/q$ by a 55$^{\circ}$ dipole magnet. After that, the continuous beam was injected into the helium buffer gas-filled radio-frequency quadrupole cooler-buncher \cite{Nieminen2001}. Finally, from there the bunched beam was sent into the JYFLTRAP double Penning trap mass spectrometer \cite{Eronen2012}.  

In the first (purification) trap of JYFLTRAP, the delivered ions were cooled, purified and centered using a mass-selective buffer gas cooling technique \cite{Savard1991}. This resulted in the extraction of only the ion of interest while removing the majority of isobaric contaminants. For $^{118}$Cd, the Ramsey cleaning method \cite{Eronen2008} with Ramsey excitation patterns (On-Off-On) 5-120-5 ms was used to remove possible contamination of $^{118}$In ions. The purified ions of interest were sent to the second (measurement) trap where their mass-over-charge ratio $m/q$ was determined using the phase-imaging ion cyclotron resonance (PI-ICR) technique \cite{Eliseev2013,Eliseev2014,Nesterenko2018,Nesterenko2021} by measuring their cyclotron frequency $\nu_c = qB/(2 \pi m)$ in a magnetic field $B$. 

\begin{figure}[h!t!b]
\includegraphics[width=\columnwidth]{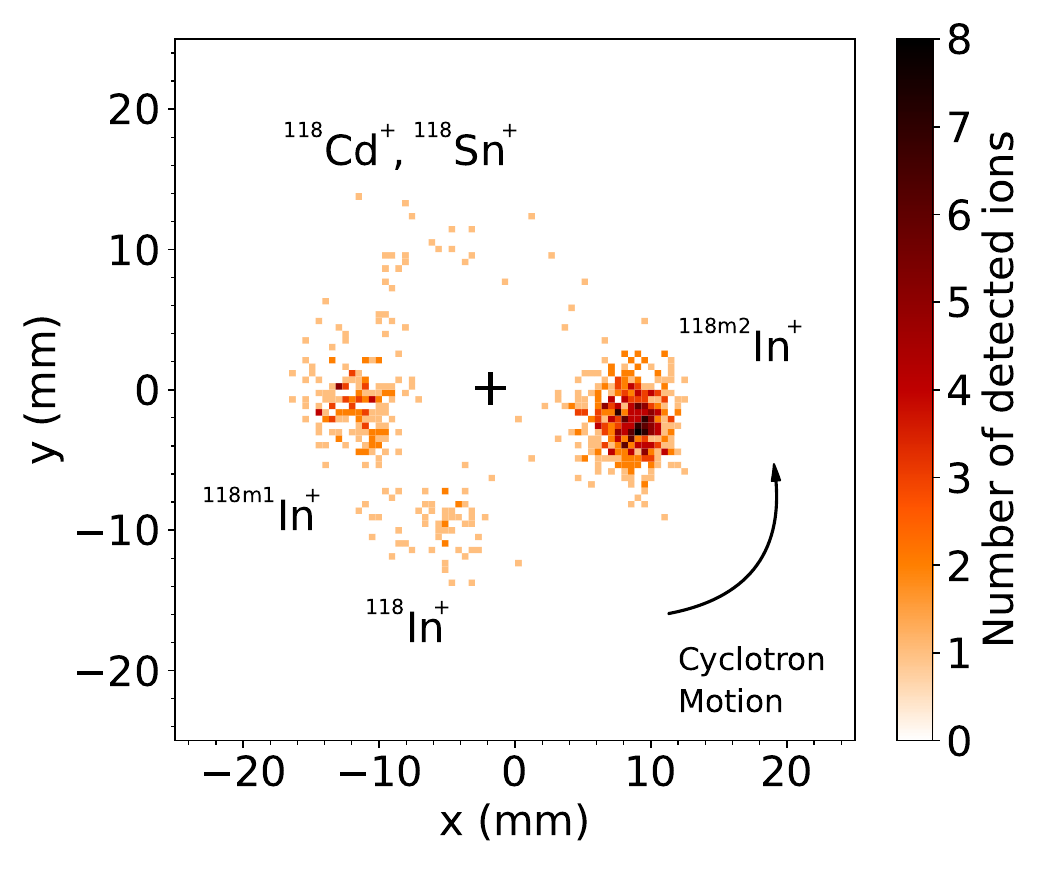}
\caption{\label{fig:118InPIICR}Projection of the cyclotron motion of $^{118}$In$^+$, $^{118m1}$In$^+$, $^{118m2}$In$^+$ states and the isobaric contaminants $^{118}$Cd$^+$ and $^{118}$Sn$^+$ ions onto the position-sensitive detector obtained with the PI-ICR technique. The phase accumulation time was set to $t_{acc} = 484$~ms and a time-of-flight gate of 64-70~$\mu$s was used. The position of the center spot is indicated with the $+$ symbol.}
\end{figure}

In the PI-ICR method, the cyclotron frequency is extracted based on the phase differences between radial motions of an ion after a phase accumulation time $t_{acc}$ (see Fig. \ref{fig:118InPIICR}). The $t_{acc}$ value was set to 800 and 408~ms for $^{118,119}$Cd and 458, 484 and 363~ms for $^{117,118,119}$In, respectively. These times were chosen to avoid an overlap between the projections of the ion of interest and possible isobaric contaminants or an isomeric state. The magnetic field $B$ was determined precisely by measuring a cyclotron frequency $\nu_{c,ref}$ of reference ions, either $^{133}$Cs$^+$, delivered from the IGISOL offline surface ion source \cite{Vilen2020} and whose mass-excess value ${\Delta_{lit.} = -88070.943(8)}$~keV \cite{AME2020} is known with high precision, or isobaric species produced together with the ion of interest.

The atomic mass $M$ is connected to the frequency ratio $r=\nu_{c,ref.}/\nu_{c}$ between the singly-charged reference ions and the ions of interest:
\begin{equation}
M = r(M_{ref}-m_e)+m_e \mathrm{,}
\end{equation}
where $m_e$ and $M_{ref}$ are the mass of a free electron and the atomic mass of the reference, respectively. For cases where the isobaric species were used as a reference, the energy difference between them (the $Q$ value) was extracted as follows:
\begin{equation}
Q = (r-1)[M_{ref} - m_e]c^2 \mathrm{,}
\end{equation} 
where $c$ is the speed of light in vacuum. 

Contribution from electron binding energies are on the order of a few eV and have thus been neglected. The measurements of the ion of interest and the reference ion were alternated to account for the temporal magnetic field fluctuations. To account for possible ion-ion interactions, in the case of $^{117}$In$^+$ and $^{118}$Cd$^+$ mass measurements a count-rate class analysis was performed \cite{Kellerbauer2003,Roux2013,Nesterenko2021}, while for the cases with significantly lower statistics the count rate was limited to one detected ion per bunch. The temporal magnetic field fluctuation of $\delta B/B = 2.01(25) \times 10^{-12}$ min$^{-1}$ $\times \delta t$, where $\delta t$ is the time between the measurements, and systematic uncertainties due to the magnetron phase advancement and the angle error were taken into account in the analysis \cite{Nesterenko2021}. For cases measured against the $^{133}$Cs$^+$ reference ions, a mass-dependent uncertainty of $\delta_m r/r = -2.35(81) \times 10^{-10} / \textnormal{u} \times (M_{ref} - M)$ and a residual systematic uncertainty of $\delta_{res}r/r=9\times 10^{-9}$ were also added \cite{Nesterenko2021}. A detailed description of systematic effects in JYFLTRAP can be found in Ref. \cite{Nesterenko2021}.

\section{Results}

\begin{table*}
\centering
\begin{ruledtabular}
\caption{\label{tab:results} Nuclides studied in this work together with their half-lives ($T_{1/2}$) and spins and parities ($J^{\pi}$) taken from the literature \cite{NUBASE2020}. The frequency ratios $r=\nu_{c,ref}/\nu_{c}$ determined using the PI-ICR technique, the reference ions (Ref.), corresponding mass-excess values $\Delta$, excitation energies $E_{x}$ and differences $\mathrm{Diff.} = \Delta - \Delta_{lit.}$ are tabulated and compared to the literature values from Refs. \cite{AME2020,NUBASE2020}. The \#~symbol denotes extrapolated mass values based on systematics.}
\begin{tabular}{lllllllllll}
Nuclide     & $T_{1/2}$  & $J^{\pi}$  & Ref. & $r=\nu_{c,ref}/\nu_{c}$  & $\Delta$ (keV)     & $\Delta_{lit.}$ (keV) & $E_x$ (keV) & $E_{x,lit.}$ (keV) & Diff. (keV) \\\hline
$^{118}$Cd 	    & 50.3(2) m  & $0^+$      & $^{133}$Cs & 	\num{0.887148498(16)}	                   & \num{-86690.0(20)} 	            & \num{-86702(20)}        & 	          &	& $12(20)$ \\
$^{119}$Cd 	    & 2.69(2) m  & $1/2^+$    & $^{133}$Cs & \num{0.894693884(17)}                       & \num{-84064.8(21)} 	        & \num{-83980(40)}          & 	          &	& $-85(40)$ \\
$^{119}$Cd$^m$  & 2.20(2) m	 & $11/2^-$   & $^{133}$Cs & 	\num{0.894695039(18)}	                   & \num{-83921.7(22)} 	        & \num{-83830(40)}          & 143.1(31)	         &	146.54(11) & $-91(40)$ \\
$^{117}$In      & 43.2(3) m & $9/2^+$      & $^{117}$Sn$^{m}$ &  \num{1.000010471(24)}   &  \num{-88942.9(26)} & \num{-88943(5)}         & 	          & & $0(6)$ \\
$^{118}$In      & 5.0(5) s & $1^+$      & $^{118}$Sn & 		\num{1.000039843(17)}        &  \num{-87277.1(20)}	        & \num{-87228(8)}         & 	          & & $-49(8)$  \\
$^{118}$In$^{m1}$      & 4.364(7) m & $5^+$      & $^{118}$Sn & 	\num{1.000040210(13)}     &  	\num{-87236.8(16)}   & \num{-87130(50)}\#        & 	 $40.3(25) $         & $100(50)$\# & $-107(50)$\# \\
$^{118}$In$^{m2}$  & 8.5(3) s   & $8^-$      & $^{118}$Sn & 	\num{1.000041470(13)}      &   \num{-87098.4(15)}   & \num{-86990(50)}\#        &   $178.7(25)$         &	$240(50)$\#\footnotemark[1]  & $-108(50)$\#\\
$^{119}$In      & 2.4(1) m   & $9/2^+$    & $^{133}$Cs & 	\num{0.894664421(17)}	                   & \num{-87712.2(21)} 	        & \num{-87699(7)}           & 	          & &  $-13(7)$\\
$^{119}$In$^m$  & 18.0(3) m	 & $1/2^-$    & $^{119}$In & 	\num{1.000002811(31)}	                   & \num{-87400.9(40)} 	        & \num{-87388(7)}           & 	$311.4(35)$          &	311.37(3) & $-13(8)$\\
\end{tabular}
\end{ruledtabular}
\footnotetext[1]{The energy difference between $^{118}$In$^{m1}$ and $^{118}$In$^{m2}$ is 138.29(14)~keV, see text for details.}
\end{table*}

The results obtained in this work and their comparison with the literature values \cite{AME2020,NUBASE2020} are presented in Table~\ref{tab:results}. The details are discussed in the following subsections.

\subsection{$^{118,119}$Cd}

All of the cadmium species reported in this work were measured against the $^{133}$Cs$^{+}$ reference ions. The mass-excess value of $^{118}$Cd, ${\Delta = -86690.0(20)}$~keV, is in agreement with the AME20 value (${\Delta_{lit.} = -86702(20)}$~keV \cite{AME2020}) based on the $Q$-value measurement of the $^{116}$Cd$(t,p)^{118}$Cd reaction \cite{Hinds1967}. Our new value is 10 times more precise.

At the same time, the $^{119}$Cd ground-state mass excess, ${\Delta = -84064.8(21)}$~keV, is 85 keV ($2.1\sigma$) lower than the literature value, ${\Delta_{lit.} = -83980(40)}$~keV \cite{AME2020}. A similar discrepancy of 91 keV was observed for $^{119}$Cd$^m$. The isomer excitation energy extracted in this work, ${E_x = 143.1(31)}$~keV is within 3.4~keV ($1.1\sigma$) from the precise NUBASE20 value, ${E_x^{lit.} = 146.54(11)}$~keV \cite{NUBASE2020}.

The mass of $^{119}$Cd in the AME20 evaluation is known only from a single $\beta$-decay study \cite{Aleklett1982}. This fact can explain the discrepancy with our work as this technique was shown to be much less accurate compared to the Penning-trap measurements, see e.g. Refs.~\cite{Ge2021,Ramalho2022,Gamage2022a,Gamage2022,Eronen2022,Hukkanen2023}.

\subsection{$^{117-119}$In}

The mass of the $^{117}$In ground state was measured against the isobaric $^{117}$Sn$^m$ (${\Delta_{lit.} = -90083.1(5)}$~keV \cite{AME2020}) strongly produced in the fusion-evaporation reaction. The extracted $Q$ value of 1140.2(26)~keV resulted in the mass excess of $\Delta = -88942.9(26)$~keV. It is in agreement with the AME20 value (${\Delta_{lit.} = -88943(5)}$~keV \cite{AME2020}) based mostly on the $\beta$-decay measurement \cite{McGinnis1955} but it is twice more precise.

In $^{118}$In all three states known in the literature \cite{NUBASE2020} were observed (see Fig. \ref{fig:118InPIICR}) and they were measured against the isobaric $^{118}$Sn isotope (${\Delta_{lit.} = -91652.8(5})$~keV \cite{AME2020}). The extracted $Q$ values for the ground, the first and the second isomeric states are $4375.7(19)$, $4416.0(15)$ and $4554.4(14)$~keV, respectively, and they resulted in the mass excesses of $-87277.1(20)$, $-87236.8(16)$ and $-87098.4(15)$~keV, respectively.

The energy difference between the first and the second isomeric state in $^{118}$In extracted in this work, ${Q(^{118}\mathrm{In}^{m2}) - Q(^{118}\mathrm{In}^{m1}) = 138.4(21)}$~keV, is in a perfect agreement with the energy difference between the $8^-$ and $5^+$ states in $^{118}$In, ${\Delta E = 138.29(14)}$~keV\footnote{This value is a weighted average of 138.2(5)~keV from a $\gamma$-spectroscopy study \cite{Hattula1969} and 138.30(15)~keV from a conversion-electron study \cite{Rissanen2007}.}. Based on this fact, we assign the first and the second isomers as the $5^+$ and the $8^-$ states, respectively, while the $^{118}$In ground state is the $1^+$ state. This level ordering is in agreement with the one proposed in the NUBASE20 evaluation \cite{NUBASE2020}.

The ground state mass excess of $^{118}$In from this work, ${\Delta = -87277.1(20)}$~keV, is $49(8)$ keV (6.1$\sigma$) lower than the literature value (${\Delta_{lit.} = -87228(8)}$~keV \cite{AME2020}) and it is four times more precise. The $Q_\beta$ value of the first isomeric state was known from the $\beta$-decay measurement (${Q_\beta = 4300(100)}$~keV \cite{Kantele1964}), however, this result was deemed irregular by the AME20 evaluators and was not used for the mass determination \cite{Huang2021}. In this work, the mass-excess value of the $^{118}$In first isomeric state was determined directly for the first time and it is $107(50)$~keV lower than the NUBASE20 extrapolation \cite{NUBASE2020} while that of the second isomeric state is $108(50)$ keV lower. It should be noted that the first-isomer mass excess, ${\Delta = -87236.8(16)}$~keV, is in agreement with the literature value for the ground-state mass excess, ${\Delta_{lit.} = -87228(8)}$~keV \cite{AME2020}.

The mass-excess value of $^{119}$In, $\Delta = -87712.2(21)$~keV, was measured against $^{133}$Cs. It differs from AME20 by 13(7) keV ($1.9\sigma$) and it is three times more precise. The isomer excitation energy extracted in this work, ${E_x = 311.4(35)}$~keV, is in agreement with the more precise NUBASE20 value ($E_x^{lit.} = 311.37(3)$~keV \cite{NUBASE2020}).

The mass of $^{118}$In was determined exclusively from a single transfer reaction study \cite{Huang2021} reported in Ref. \cite{Pilt1985}. The results from this publication have also the biggest weight (86\% \cite{Huang2021}) in the $^{119}$In mass determination. The remaining 13\% is based on the $^{120}$Sn$(d,^{3}$He$)^{119}$In study reported in Ref. \cite{Weiffenbach1971} while the $\beta$-decay measurement \cite{Aleklett1982} has a weight of about 1\%. Our results allow us to conclude that the calibration used in Ref. \cite{Pilt1985} was most likely not correct. It should be noted that since the mass of $^{119}$Cd is partially based on the $Q_\beta(^{119}\mathrm{In})$ value \cite{Huang2021}, our updated $^{119}$In mass value also affects the mass of $^{119}$Cd. 

\section{Discussion}

To analyze the influence of the updated mass-excess values, the two-neutron separation energies $S_{2n}$ were calculated and compared to the AME20 \cite{AME2020} values. They are defined as:
\begin{equation}
S_{2n}(Z,N) = \Delta(Z,N-2) - \Delta(Z,N) +2\Delta_n \mathrm{,}
\end{equation}
where $\Delta(Z,N)$ is a mass excess of a nucleus with given proton ($Z$) and neutron ($N$) numbers and $\Delta_n$ is the mass excess of a free neutron. In addition, the two-neutron shell-gap energies $\delta_{2n}$ which show the differences between the $S_{2n}$ values and are defined as: 
\begin{equation}
\delta_{2n}(Z,N) = S_{2n}(Z,N) - S_{2n}(Z,N+2) \mathrm{,}
\end{equation}
are also extracted. For the indium isotopic chain, we have also included the recent JYFLTRAP results for $^{120-124}$In reported by Nesterenko \textit{et al.} \cite{Ruotsalainen2023} as they were published after the AME20 evaluation cut-off date \cite{AME2020}.

\begin{figure}[h!t!b]
\includegraphics[width=\columnwidth]{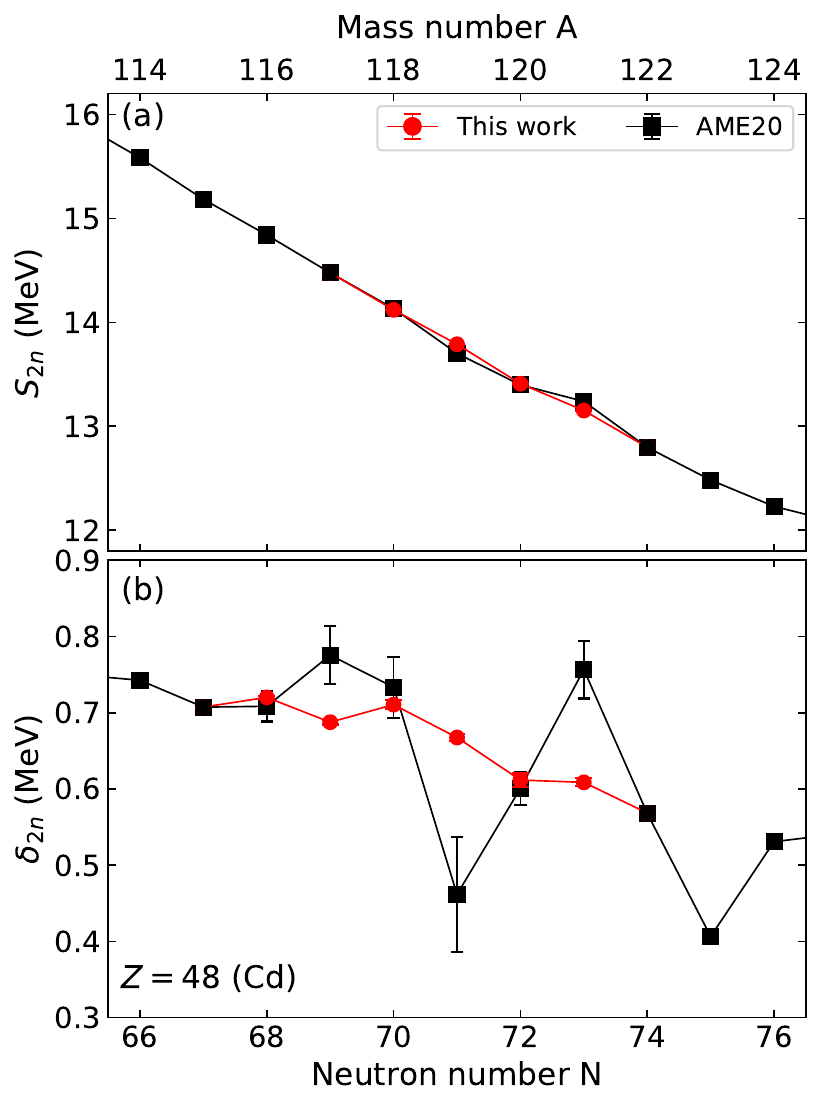}
\caption{\label{fig:cdmasstrends}A comparison of a) two-neutron separation energies $S_{2n}$ and b) two-neutron shell-gap energies $\delta_{2n}$ for the Cd isotopic chain between AME20 \cite{AME2020} and the results from this work.}
\end{figure}

In the cadmium isotopic chain a significant staggering of the $S_{2n}$ values observed around $A=119$ ($N=71$) disappears when the mass-excess values reported in this work are used, see Fig. \ref{fig:cdmasstrends}a. One can notice that the new values are smoothing the trend while removing the small deviations at $A=119$ ($N=71$) and $A=121$ ($N=73$). This effect is even more visible when analyzing the $\delta_{2n}$ values, see Fig. \ref{fig:cdmasstrends}b.

\begin{figure}[h!t!b]
\includegraphics[width=\columnwidth]{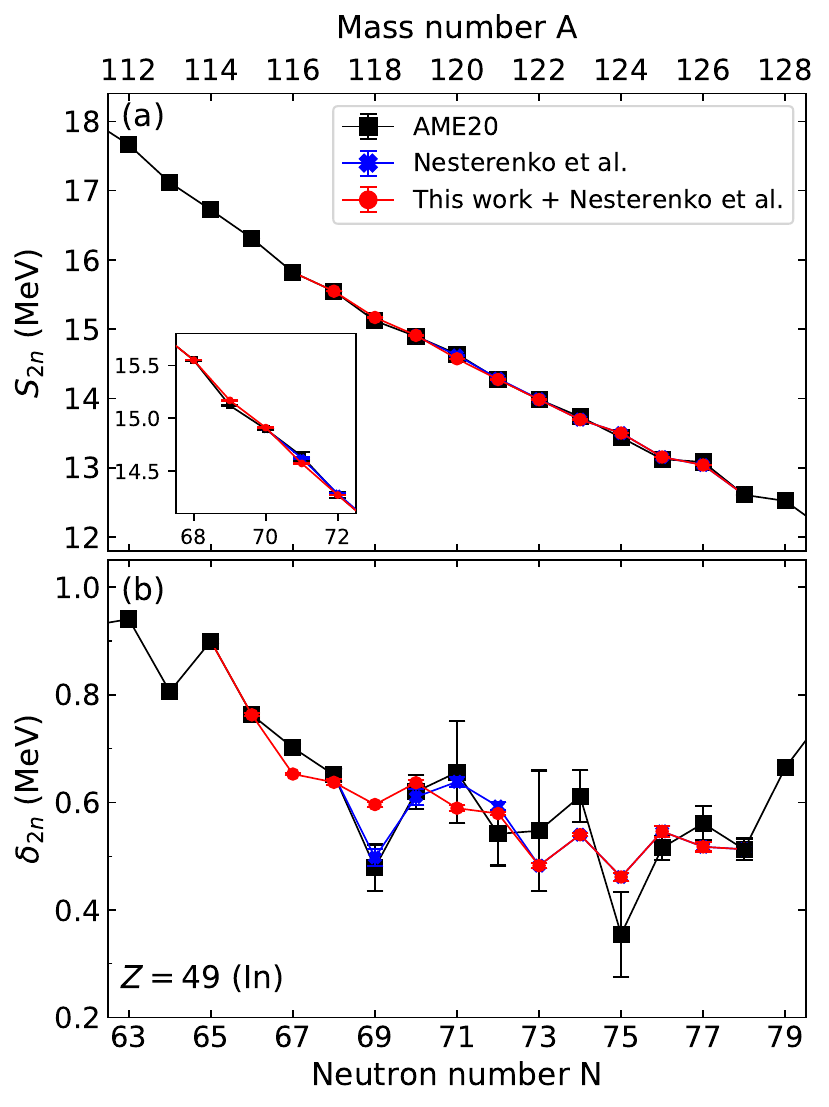}
\caption{\label{fig:inmasstrends}A comparison of a) two-neutron separation energies $S_{2n}$ and b) two-neutron shell-gap energies $\delta_{2n}$ for the In isotopic chain between AME20 \cite{AME2020}, the work by Nesterenko \textit{et al.} \cite{Ruotsalainen2023}, where mass values of $^{120-124}$In are reported, and the combined results from JYFLTRAP (this work and Nesterenko \textit{et al.} \cite{Ruotsalainen2023}). The close-up of the $S_{2n}$ curve around $N=70$ is plotted in the inset.}
\end{figure}

In the case of the indium isotopic chain, the biggest differences in the $S_{2n}$ chain are observed for $A=118$ ($N=69$) and $A=120$ ($N=71$),  where the updated mass values are leading to a more linear trend, see Fig.~\ref{fig:inmasstrends}a. The influence of the results from this work is better visible in the $\delta_{2n}$ plot (see Fig.~\ref{fig:inmasstrends}b). In particular, a dip in the trend at $A=118$ ($N=69$) almost disappears, making the $\delta_{2n}$ curve smoother. 

Changes in the mass trends can be interpreted as an indication of a (sub)shell closure or an onset of nuclear collectivity \cite{Lunney2003}. However, for isotopic chains considered in this work, our new mass measurements result in a clear smoothing of their respective $S_{2n}$ trends. As a result, we must come to the conclusion that the previously observed irregularities in the $S_{2n}$ trends were not due to a change in the structure of the nuclei in the region but rather arose from imprecise mass excess values extracted from $\beta$-decay and reaction studies. This conclusion is consistent with the laser-spectroscopy studies of charge radii and electromagnetic moments \cite{Yordanov2013,Yordanov2016,Vernon2019} where no significant deviations were reported around that mass region for neither cadmium nor indium isotopes. 

\begin{table}
\centering\small
\begin{ruledtabular}
\caption{\label{tab:logft}A comparison of the $\log(ft)$ values for the $^{A}\mathrm{Cd}(0^+_{gs})\longrightarrow$ $^{A}\mathrm{In}(1^+_1)$ decay calculated with the $\log(ft)$ calculator \cite{logft} using the $Q_\beta$ and the half-lives $T_{1/2}$ values from AME20/NUBASE20 \cite{AME2020,NUBASE2020} with the results from JYFLTRAP (this work and Ref. \cite{Ruotsalainen2023}).}
\begin{tabular}{cllll}
 & $A$ & 118 & 120 & 122 \\\hline
 & $Q_\beta$ (keV) & 527(21) &  1770(40) & 2960(50) \\
AME20& $T_{1/2}$ & 50.3(2) m & 50.80(21) s & 5.24(3) s \\
& $\log(ft)$ &  3.93(6) & 4.09(4) & 4.02(4) \\\hline
 & $Q_\beta$ (keV) & 587.1(28) &  1752.1(46) &  2861.1(25) \\
JYFLTRAP & $T_{1/2}$ & 50.3(2) m & 50.80(21) s & 5.98(10) s \\
& $\log(ft)$ & 4.089(8) & 4.077(5) & 4.019(8) \\
\end{tabular}
\end{ruledtabular}
\end{table}

To further test the hypothesis of an absence of a structure change in the $N\approx70$ mass region, the $\log(ft)$ values between the parents, even-even cadmium isotopes, and the daughters, the odd-odd indium isotopes, were calculated for $^{118,120,122}$Cd isotopes. These values provide direct access to the structure of the initial and the final states as they are linked to the reduced transition probabilities. The $\log(ft)$ values can be extracted when the $Q_\beta$, half-life and the branching ratio values to a given state are known. 

The updated masses of $A=118$ isotopes from this work are leading to a new $Q_\beta$ value for $^{118}$Cd and, consequently, a re-evaluated $\log(ft)$ value. For comparison, the $\log(ft)$ was also extracted for $^{120,122}$Cd using the new mass-excess values and a new half-life for $^{122}$Cd reported in Ref.~\cite{Ruotsalainen2023}. These three isotopes are particularly interesting as they are decaying exclusively to the long-lived $1^+_1$ states in the In isotopes \cite{ENSDF}, allowing, on the one hand, an extraction of the $\log(ft)$ values with high precision and, on the other hand, to study pure Gamow-Teller $\beta$ decays. The comparison between the results from the AME20/NUBASE20 evaluations \cite{AME2020,NUBASE2020} and JYFLTRAP (this work and Ref.~\cite{Ruotsalainen2023}) is presented in Tab. \ref{tab:logft}.

For $^{120,122}$Cd the JYFLTRAP $\log(ft)$ values are consistent with the AME20/NUBASE20 results, however, they have a higher precision. It is worth noting that in the case of $^{122}$Cd, a decrease of the $Q_\beta$ value by about 100~keV ($\approx$2$\sigma$) was compensated by a significant increase of the half-life, from 5.24(3)~s to 5.98(10)~s, as reported in Ref.~\cite{Ruotsalainen2023}. The mass measurements of $^{118}$Cd and $^{118}$In reported in this work resulted in an increase of $Q_\beta$ from 527(21)~keV \cite{AME2020} to 587.1(28)~keV. This has lead to a significant change of the $\log(ft)$ value for $^{118}$Cd, from 3.93(6) to 4.089(8), which is now in line with the $^{120,122}$Cd cases.

Our analysis indicates that with an increasing number of neutrons the structure of the cadmium and indium isotopes is not changing significantly. In addition, it can be deduced that the main components of the wave functions in both isotopic chains remain dominant as the extracted $\log(ft)$ values ($\approx4.05$) are relatively low even for the allowed $0^+\rightarrow 1^+$ decays, see Fig.~9a in Ref.~\cite{Turkat2023}. 

\section{Conclusions}

The masses of neutron-rich $^{118,119}$Cd and $^{117-119}$In isotopes were measured for the first time using the JYFLTRAP double Penning trap. The extracted mass-excess values for $^{118}$Cd and $^{117}$In are in agreement with AME20 \cite{AME2020}, however, they are ten and two times more precise, respectively. In case of $^{119}$Cd and $^{118,119}$In, the measured masses differ from the evaluation \cite{AME2020} by 85~keV (6.1$\sigma$), 49~keV (1.9$\sigma$) and 13~keV (2.1$\sigma$), respectively. 

The excitation energy of the $^{119}$In isomer is in agreement with the more precise NUBASE20 value \cite{NUBASE2020} while for $^{119}$Cd$^{m}$ it differs by 1.1$\sigma$ (3.4 keV). All three long-lived states in $^{118}$In were observed and the isomer excitation energies were extracted for the first time. The energy difference between $^{118}$In$^{m2}$ and $^{118}$In$^{m1}$, ${\Delta E = 138.4(21)}$~keV, extracted in this work is in agreement with the literature energy difference between the $8^-$ and $5^+$ states, ${\Delta E = 138.29(14)}$~keV \cite{Hattula1969,Rissanen2007}. This fact was used for the identification of the measured states. 

The updated mass values have removed the irregularities observed earlier in the two-neutron separation energies and two-neutron shell gap curves for both isotopic chains. In particular, the staggering observed around $N=70$ has disappeared. The re-evaluated $\log(ft)$ value for the $^{118}$Cd decay increased from 3.93(6) to 4.089(8) and it is now similar to the two more neutron-rich $^{120,122}$Cd isotopes. 

Our results indicate that there are no significant structural changes, such as a subshell closure or a change of deformation, in neither cadmium nor indium isotopes at around $N=70$. They also show the importance of measuring radioactive species close to stability as the previously used experimental methods were often lacking the required accuracy. 

\begin{acknowledgments}

This project has received funding from the European Union’s Horizon 2020 research and innovation programme under grant agreements No. 771036 (ERC CoG MAIDEN) and No. 861198–LISA–H2020-MSCA-ITN-2019 and from the Academy of Finland projects No. 295207, 306980, 327629, 354589 and 354968. J.R. acknowledges financial support from the Vilho, Yrj\"o and Kalle V\"ais\"al\"a Foundation. D.Ku. acknowledges the support from DAAD grant number 57610603.

\end{acknowledgments}

\bibliographystyle{apsrev}
\bibliography{mybibfile}

\end{document}